\documentclass[12pt]{article}

\usepackage[
colorlinks=true,backref,pagebackref]{hyperref} 

\usepackage{epsfig}
\usepackage{amsbsy}
\usepackage{times}
\usepackage{amsmath}
\usepackage{amsfonts}
\usepackage{amssymb}
\usepackage{multicol}
\usepackage{color}
\usepackage{graphicx}
\usepackage{pst-coil}
\usepackage{colortbl}

 \usepackage{textcomp}

\definecolor{dark-green}{rgb}{0,0.7,0}
\definecolor{dark-blue}{rgb}{0,0.2,0.5}
\definecolor{med-blue}{rgb}{0,0.7,1}
\definecolor{mblue}{rgb}{0,0.2,1}
\definecolor{cnc}{rgb}{0.8,0,0}
\definecolor{light-red}{rgb}{1,0.8,0.8}
\definecolor{dark-yellow}{rgb}{1,0.8,0}
\definecolor{light-blue}{rgb}{0.8,0.9,1}
\definecolor{verylight-blue}{rgb}{0.93,0.95,1}
\definecolor{light-yellow}{rgb}{1,0.9,0.8}
\definecolor{grey}{gray}{0.88}

\def\a{\alpha}
\def\b{\beta}
\def\g{\gamma}
\def\d{\delta}

\def\r{\rho}
\def\vta{\vartheta}
\def\vt{\vartheta}
\def\stareq{\stackrel{*}{=}}

\def\ltextindent#1{\hbox to \hangindent{#1\hss}\ignorespaces}

\def\negenspace{\kern-1.1em}

\def\sqr#1#2{{\vcenter{\hrule height.#2pt\hbox{\vrule width.#2pt
height#1pt \kern#1pt \vrule width.#2pt}\hrule height.#2pt}}}

\begin{document}

\title{Gauge Theory of Gravity and Spacetime\footnote{Based on an
    invited seminar, given at: {\it Towards a Theory of Spacetime
      Theories}, International Workshop, 21 to 23 July 2010, IZWT,
    Bergische Universit\"at Wuppertal, Germany; Dennis~Lehmkuhl,
    Erhard~Scholz, and Gregor~Schiemann (organizers). To be published
in {\it Einstein Studies} (Birkh\"auser, Boston, MA, 2015).}}


\author{Friedrich W.\ Hehl\\Univ.\ of Cologne and Univ.\ of
  Missouri,~Columbia,~MO\\hehl@thp.uni-koeln.de} \date{05 May
  2014$\quad${\it file WuppertalWorkshop04.tex}}
\maketitle

\begin{abstract}

  \noindent The advent of general relativity in 1915/16 induced a
  paradigm shift: since then, the theory of gravity had to be seen in
  the context of the geometry of spacetime.  An outgrowth of this new
  way of looking at gravity is the gauge principle of Weyl (1929) and
  Yang--Mills--Utiyama (1954/56). It became manifest around the 1960s
  (Sciama--Kibble) that gravity is closely related to the Poincar\'e
  group acting in Minkowski space. The gauging of this external group
  induces a Riemann--Cartan geometry on spacetime. If one generalizes
  the gauge group of gravity, one discovers still more involved
  spacetime geometries. If one specializes it to the translation
  group, one finds a specific Riemann--Cartan geometry with
  teleparallelism (Weitzenb\"ock geometry).
\end{abstract}
\newpage

\tableofcontents
\bigskip

\section{Apropos a theory of spacetime theories}

In this workshop, we are supposed to move ``Towards a theory
of spacetime theories''. The idea seems to be that there are many
spacetime theories around, the Riemannian spacetime theory in the
framework of general relativity (GR), the Weitzenb\"ock spacetime
theory in teleparallelism approaches to gravity, the Riemann--Cartan
spacetime theory withing the Poincar\'e gauge theory of gravity (PG),
the superspace(time) theory within supergravity, the Weyl(--Cartan)
spacetime theory within a gauge theory of the Weyl group, etc.. The
list could be continued with spacetime theories emerging in
quantization approaches to gravity where spacetime becomes mostly a
discrete structure. There is a plethora of different spacetime
theories around and it is hardly possible to view all of them {}from
some kind of a unifying principle, let alone {}from one theory
encompassing these spacetime theories as specific subcases.

Orientation in this seemingly chaotic landscape of spacetime theories
can be provided by looking at the successful theories of our days that
are able to predict and describe correctly fundamental phenomena
occurring in nature. There is the standard model of particle physics,
based on the Poincar\'e group (also known as inhomogeneous Lorentz
group) and the internal groups $SU(3), SU(2), U(1)$. The Poincar\'e
group is the group of motion in the Minkowski spacetime of special
relativity (SR) and it classifies the particles according to their
masses and their spins. The internal groups describe the strong and
the electro-weak interactions by means of the respective gauge (or
Yang--Mills) theory.

A book on the centennial of the discovery of SR was called
\cite{EhlersClaus}: ``Special Relativity. Will it survive the next 100
years?'' When I read this title in 2005, I thought for a moment that I
must have been in a time machine and in reality I am living in
1905. Hadn't SR already been superseded in 1915/16 by GR, I wondered?
I pointed this out to the editors that this title looks anachronistic
to me and is hardly appropriate for editors who both are known to
subscribe to GR. It turned out that both wanted to ask whether SR
survives {\it locally} as a valid theory. But they didn't want to
change the title since this fact was, as they told me, known to
everybody anyway. I gave up since I realized that in a time when in
the tabloid press a title is more for catching one's attention than
for spreading the truth, the scientific literature cannot stand aside.

But what is my point? Well, we all seem to agree that at least
presently SR is universally valid {locally} {\it in a freely falling
  frame.} So far no deviations therefrom have been found. Only at very
high accelerations, the principle of locality, inherent in SR, may
need to be amended \cite{Bahram}. In any case, our march towards a
theory of all spacetime theories has at least a definitive starting
point.

But was SR superseded by GR? Yes, of course---in spite of the title of
reference \cite{EhlersClaus}. The abstraction of a Minkowski space can
only be uphold when gravitational effect can safely be neglected. If
you measure Planck's constant or the elementary charge by a
conventional laboratory experiment, then this assumption is
justified. But if you go down the stairs, you had better not neglect
gravity, otherwise you may fall downwards; or if you measure the
deflection angle of a light ray gracing a star, you also better don't
neglect gravity. {}From the laboratory to at least the scale of the
planetary system, GR is in excellent agreement with experiment. On the
galactic scale this is taken for granted by most physicists, but this
is disputed by supporters of MOND, of TeVeS, of f(R)-theory, or of
nonlocal gravity,\footnote{Mashhoon and the author
  \cite{Hehl:2008eu,Hehl:2009es} formulated a {\it nonlocal}
  translational {\it gauge theory} of gravity that seems to be able to
  reproduce the observed rotation curves of galaxies, see the most
  recent results in \cite{Rahvar:2014yta}.} for example, compare the
presentations in \cite{Famaey}. Anyway, GR is mostly accepted for the
global description of the cosmos and if the cosmological principle is
assumed, namely homogeneity and isotropy of space, Einstein's field
equation predicts a Friedmann cosmos. The cosmos started with the Big
Bang and it is usually assumed to be equipped with a scalar
inflationary field providing a sufficiently fast expansion. Needless
to say that this framework is based on a number of extreme
extrapolations.

The message is then that the Minkowski spacetime picture is
substituted by the Riemannian one. But this doesn't rest on the same
strong experimentally well-confirmed basis as the local presence of
the Minkowski spacetime of SR.

\section{Is the gauge idea the underlying principle for all
  interactions?}

Since the advent of GR it was clear that a spacetime theory is
inextricably linked to gravity. One cannot be understood without the
other. Coming back to the topic of our workshop, it is then clear that
gravity has to be considered in this general context willy
nilly. Accordingly, a spacetime theory is at the same time, at least
in some of its parts, a theory of gravity.

Let us then turn to gravity: Is GR all we have? Well, by some people
GR is declared to be sacrosanct and you may touch it only by
superimposing some abstract mathematical framework supposedly
quantizing GR, see \cite{Kiefer}. But practitioners of this method
increasingly become aware that they have to amend the
Hilbert--Einstein Lagrangian of the free gravitational field by
non-Riemannian supplementary terms thereby dissolving to a certain
extend the Riemannian structure they started with
\cite{Diakonov:2011fs,Baekler:2011jt,Obukhov:2012je}. Hence
alternatives to GR gain credibility even if GR is left fixed at first.

Is GR the only reasonable theory of gravity? No, it isn't. Already in
1956 Utiyama began to formulate gravity as a gauge theory, for a
selection of classical papers, see \cite{Reader}. The strong and
electro-weak gauge theories are based on internal symmetry
groups---mathematically semi-simple Lie groups---linked to conserved
currents. The gauge idea basically requires that the {\it rigid} (or
global) symmetry group related to the conserved current under
consideration has to be made {\it local}; without giving up the
invariance of the Lagrangian, this is only possible by the
introduction of a gauge potential $A=A_i dx^i$ (a covector or an
1-form) that transform under this group suitably; for each parameter
of the group one needs one covector field. Thus, the group dictates
the interaction emerging {}from that scheme: a new interaction is
created {}from a conserved current via the (reciprocal) Noether
theorem and the symmetry group attached to it.

In the standard model of particle physics all gauge groups are
internal, that is, they act in some internal space. In the original
Yang--Mills theory, for example, it was the isospin space. But the
gauge idea of {\it localizing a symmetry} does not seem to be
restricted to internal groups. An external group affects by definition
spacetime.  If we have a conserved current and a corresponding group,
nothing prohibits us to apply the gauge principle.

How does gravity come into this framework? The source of Newtonian
gravity is the mass of a body. In classical physics, mass is a
conserved quantity, as has been experimentally demonstrated by
Lavoisier (around 1790). In SR mass conservation is no longer
valid---as has been shown in the 1930s by more accurate experimental
techniques---and is superseded by energy-momentum conservation, as has
been most vividly demonstrated in Alamogordo in 1945. Clearly then,
the Poisson equation controlling Newton's gravitational potential
$\phi$, namely $\Delta \phi(\mathbf{r},t) = 4\pi G
\rho(\mathbf{r},t)$, with $\Delta$ as the Laplace operator, $G$ as the
gravitational constant, and $\rho$ as the mass density, has to be
substituted by an equation that carries on its right-hand-side the
energy density of matter (and/or radiation). However, according to SR,
the energy density is the time-time component of the symmetric
energy-momentum current $\mathfrak{t}_{ij}=\mathfrak{t}_{ji}$ of
matter (and/or radiation).

For an isolated physical system, the energy-momentum current
$\mathfrak{t}_{ij}$ is conserved: $\partial_j\mathfrak{t}_{i}{}^j
=0$. This is an expression of the fact that the action of the system
is invariant under translations in time and space. Consequently, the
conserved energy-momentum current together with the translation group
$T(4)$ acting in Minkowski space should underlie gravity. Since the
translation group has four parameters, one describing a time
translation and three describing space translations, we expect four
potential one-forms $\vt^\a$, for $\a=0,1,2,3$. As we will see further
down, this framework leads to a teleparallelism theory of gravity
and back to a theory that is equivalent to GR for conventional (bosonic)
matter. Accordingly, GR can be understood as a gauge theory of the
translation group $T(4)$, which is an {\it external} group.

Ergo, all interactions, including gravity, are governed by gauge field
theories. But let us now turn back to the history of the gauge idea:

\section{The gauging of the Poincar\'e group}

As we mentioned before, Utiyama \cite{Utiyama} first attacked the
problem of understanding gravity as a gauge theory by means of gauging
the Lorentz group $SO(1,3)$. In this way, Utiyama supposedly derived
general relativity. However, the problematic character of his
derivation is apparent. First of all, he had to introduce in an ad hoc
way tetrads $e_i{}^\alpha$ (or coframes $\vt^\a=e_i{}^\a dx^i$), first
holonomic (natural) and later anholonomic (arbitrary)
ones. Secondly, he has to assume the connection
$\Gamma_i{}^{\alpha\beta}$ of spacetime to be Riemannian, without any
convincing argument. 

But thirdly, perhaps the strongest reason, the current linked to the
(homogeneous) Lorentz group is the {\it angular momentum current}
$\frak{J}_{ij}{}^k=-\frak{J}_{ji}{}^k$, which is conserved,
$\partial_k\frak{J}_{ij}{}^k=0$. However, as we have seen in the last
section, gravity is coupled to the conserved and symmetric
energy-momentum current $\frak{t}_{ik}$.  Accordingly, Einstein (1915)
took in general relativity the symmetric energy-momentum current
$\frak{t}_{ik}$ as the source of gravity in his field equation and
{\it not} the angular momentum current.  Hence Utiyama was not on the
right track. Interestingly enough, in numerous publications even
today, the Lorentz group is incorrectly thought of as gauge group of
GR; usually the conserved current coupled to it is not even mentioned.

This can be also viewed {}from the translational gauge group of
gravity, at which we arrived above. In a Minkowski space, as in any
Euclidean space, the group of motions consists of translations {\it
  and} rotations. In fact, the semidirect product of the translation
group and the Lorentz group, $T(4)\!\rtimes\!  SO(1,3)$, is the
Poincar\'e group $P(1,3)$ with its $4+6$ parameters (and its $4+6$
gauge potentials $\vt^\a$ and $\Gamma^{\a\b}=-\Gamma^{\b\a}$,
respectively). In a Euclidean or Minkowskian space the translations do
not live alone, they are accompanied, in a nontrivial way, by the
(Lorentz) rotations. Accordingly, since we find reasons to gauge the
translations in a Minkowski spacetime, it is hardly avoidable to gauge
also the rotations. If one has spinless matter, this argument may be
skipped. However, if we have fermionic matter, its rotational behavior
is closely linked to the translational behavior. Kibble, who was the
first to gauge the Poincar\'e group \cite{Kibble:1961}, poses the
following question \cite{Kibble}:
\begin{quotation} ``... Is it possible that starting {}from a theory with
    rigid symmetries and applying the gauge principle, we can recover
    the gravitational field?  The answer turned out to be yes, though
    in a subtly different way and with an intriguing twist.  Starting
    {}from special relativity and applying the gauge principle to its
    Poincar\'e-group symmetries leads most directly not precisely to
    Einstein's general relativity, but to a variant, originally
    proposed by \'Elie Cartan, which instead of a pure Riemannian
    space-time uses a space-time with torsion.  In general relativity,
    curvature is sourced by energy and momentum.  In the Poincar\'e
    gauge theory, in its basic version, additionally torsion is
    sourced by spin.''
\end{quotation}

This is also the basic message of our seminar: Gauging an external
group, here the Poincar\'e group, leads directly to a new geometry of
spacetime, here the Riemann--Cartan geometry of spacetime. To an
external gauge group a certain geometry of spacetime is attached, the
Minkowski space is deformed in accordance with the gauged
symmetries. Moreover, without a conserved current, there can be no
real gauge procedure in the sense of Weyl and Yang--Mills. If somebody
tries to sell you a gauge theory without mentioning the associated
conserved current, don't believe her or him a word. Gauging the Weyl
group {\it without} considering the scale current and gauging the
conformal group {\it without} considering the conformal currents, are
procedures that may lead to something, but certainly not to gauge
theories \`a la Weyl--Yang--Mills, see the discussion in \cite{Reader}.

Often I have heard the argument that gravity can have no relation to
the translation group since GR takes place in a Riemannian space and
therein the translations are an ill-defined concept since they are not
integrable, for example. However, this argument rests on a
misunderstanding. In a gauge approach, at the start of the procedure,
that is, before the rigid symmetry is made local, we consider the
gravity-free case. Accordingly, we are in Minkowski space where a
translation is part of the group of motion. Only after we localized
the symmetry, we lose the underlying Minkowski space, it gets
deformed, and one has to reconstruct the emerging geometry. This is
the radicality of the gauge principle: an interaction is created by a
symmetry. The translation group $T(4)$, a subgroup of the Poincar\'e
group $P(1,3)$, which acts in a Minkowski space, creates the
gravitational potential $\vt^\a$. The Lorentz subgroup $SO(1,3)$
creates another gravitational potential
$\Gamma^{\a\b}=-\Gamma^{\b\a}$, the consequences of which we will have
to discuss.

\section{Einstein's discussion of the transition {}from special to
  general relativity}

Before we turn to the subject of the gauging of the Poincar\'e group,
we remind ourselves how Einstein ``derived'' gravity
\cite{Meaning}. When Einstein developed GR, he could take a classical
mass point with mass $m$ as a starting point for his
investigations. He studied its behavior in an accelerated reference
system. Technically, in order to switch on acceleration, he
transformed the original Cartesian coordinate system $X^i$ to a
curvilinear coordinate system $x^{i}$. Let us look at this in more
detail. The points in the Minkowski space of SR can be described with
the help of Cartesian coordinates $X^i$, with $i=0,1,2,3$. In these
coordinates, the line element reads
\begin{equation}\label{lineelement}
  ds^2=(dX^0)^2-(dX^1)^2-(dX^2)^2-(dX^3)^2=o_{ij}dX^i\otimes dX^j,
\end{equation}
with $o_{ij}=\text{diag}(1,-1,-1,-1)$ and summation over repeated
indices. The equation of motion of a {\it force-free} mass in an
inertial frame $K$,
\begin{equation}\label{eqmo1}
\frac{d^2X^k}{ds^2}=0,
\end{equation}
leads for the particle trajectory to a straight line with constant
velocity.

The same motion, as viewed {}from the accelerated frame $K'$, can be
derived by a transformation of (\ref{eqmo1}) to curvilinear
coordinates,
\begin{equation}\label{eqmo2}
  \frac{D^2x^k}{Ds^2}:=\frac{d^2x^k}{ds^2}
  +\widetilde{\Gamma}_{ij}{}^k\frac{dx^i}{ds}\frac{dx^j}{ds}=0,
\end{equation}
with the Riemannian connection (Christoffel symbols of the 2nd kind):
\begin{equation}\label{Christ}
  \widetilde{\Gamma}_{ij}{}^k:={\scriptstyle\frac 12}
  g^{k\ell}\left(\partial_i g_{j\ell}-\partial_j g_{i\ell}
    +\partial_\ell g_{ij} \right)=\widetilde{\Gamma}_{ji}{}^k;
\end{equation}
here we abbreviated the partial differentiation $\partial/\partial
x^i$ as $\partial_i$. The massive particle accelerates with respect to
the non-inertial frame $K'$ in such a way that this acceleration is
independent of its mass. But an observer in $K'$ cannot tell whether
this motion is accelerated or induced by a homogeneous gravitational
field of strength $\widetilde{\Gamma}_{ij}{}^k$. In other words, the
reference system $K'$ can be alternatively considered as being at rest
with respect to $K$, but a homogeneous gravitational field is present
that is described by the {{\em Christoffel symbols}}
$\widetilde{\Gamma}_{ij}{}^k$.

Nothing has happened so far. We are still in a Minkowski space in
which---as is shown in geometry---the {{\em Riemann
    curvature tensor}} belonging to the Christoffel symbols
\begin{equation}
  \widetilde{R}_{ijk}{}^\ell := 2\partial_{[i}
  \widetilde{\Gamma}_{j]k}{}^\ell+2
  \widetilde{\Gamma}_{[i|m|}{}^\ell \,
  \widetilde{\Gamma}_{j]k}{}^m
\end{equation}
vanishes, that is $ \widetilde{R}_{ijk}{}^\ell =0$; brackets around
indices denote antisymmetrization: $[ij]:=\{ij -ji\}/2$. This is the
ingenuity of Einstein's approach: He considers force-free motion
{}from two different reference frames and identifies thereby the
Christoffels as describing---according to the equivalence
principle---a homogeneous gravitational field. Of course, this
gravitational field in Minkowski space is fictitious, it is simulated,
it doesn't really exist since the Riemann curvature vanishes.

Besides massive point particles, we have light rays (``photons'') that
can be considered in a similar way. For light propagation we have
$ds^2=0$, but the geodesic line (\ref{eqmo2}) can be reparametrized
with the help of a suitable affine parameter. Then, {}from the point of
view of reference frame $K'$, a light ray that propagates in a
straight line in the inertial frame $K$ appears to be deflected in
$K'$. According to Einstein \cite{EinsteinReprint}, ``...the principle
of the constancy of the {{\em velocity of light}} {\it in vacuo} must
be modified, since we easily recognize that the path of the light ray
with respect to $K'$ must in general be curvilinear.'' Thus, the
gravitational field deflects light. This is one of Einstein famous
and successful predictions.

In order to create a real gravitational
field---this is Einstein's assumption---we must relax the rigidity of
Minkowski space and allow for Riemannian curvature, inducing in this
way a ``deformed'' spacetime carrying non-vanishing curvature $
\widetilde{R}_{ijk}{}^\ell \ne0$. A prerequisite for this
procedure to work is the fact that the Christoffels depend at most on
first derivatives $\partial_k g_{ij}(x)$ of the metric
$g_{ij}(x)$. These first derivatives appear even in a flat space in
an accelerated frame. Only non-vanishing second derivatives tell us
about real gravitational fields.

There is one more thing to be seen {}from (\ref{eqmo2}). If we multiply
it with a slowly varying scalar mass density $\rho$ of dust matter,
then we recognize that the Christoffels are coupled to the (symmetric)
energy-momentum tensor density of dust,\footnote{A more detailed
  discussion can be found in Adler, Bazin, and Schiffer \cite{Adler},
  p.~351.}
\begin{equation}\label{eqmo3}
  \rho\frac{d^2x^k}{ds^2}
  +\frak{t}^{ij}\,\widetilde{\Gamma}_{ij}{}^k= 0
  \qquad\text{with}\qquad \frak{t}^{ij}:=\rho
  u^i u^j
\end{equation}
and $u^i:={dx^i}/{ds}$ as velocity of the dust. The fictitious
non-tensorial force density $\frak{f}^k:=\frak{t}^{ij}\,
\widetilde{\Gamma}_{ij}{}^k$, as observed by Weyl \cite{WeylSTM}, is
somewhat analogous to the Lorentz force acting on a charged particle
in electrodynamics $\frak{f}^k_{\text{Lor}}:= \frak{J}^i F_i{}^k$,
with $\frak{J}^i=\r_{\text{el}}u^i$ as electric current density and
$F_{ik}$ as electromagnetic field strength, the difference being that
here the force density $\frak{f}^k$ is quadratic in $u^i$, whereas the
Lorentz force density $\frak{f}^k_{\text{Lor}}$ is linear in $u^i$; note
also that the electromagnetic field is antisymmetric $F_{ik}=-F_{ki}$
and the gravitational field symmetric $\widetilde{\Gamma}_{ij}{}^k=+
\widetilde{\Gamma}_{ji}{}^k$. Thus, as a byproduct, we have identified
the energy-momentum tensor density of matter as the {{source of
    gravity}}.

\section{Neutron interferometer experiments}

However, in the meantime, I mean since 1916, we have learned that
there are fermions in nature. Besides mass $m$, they carry
half-integer spin $s$. Instead of a mass point, we will then study the
simplest massive fermion, the Dirac field in an inertial and a
non-inertial reference frame thus taking care of Synge's verdict {\it
  ``Newton successfully wrote apple = moon, but you cannot write apple
  = neutron''}. This is what, in fact, Kibble \cite{Kibble:1961} has
done in 1961.

But even better, experimentally it has been clear since 1975 that the
Colella--Overhauser--Werner (COW) experiment \cite{COW} is the
``modern'' archetypal experiment for a fermion in a gravitational
field: A monochromatic neutron beam, extracted {}from a nuclear reactor,
falls freely in the gravitational field of the earth. The phase shift
of its wave function $\Psi(x)$, caused by the gravitational field, is
measured by means of an interferometer built {}from a silicon
mono-crystal, see also \cite{RauchW}. Accordingly, the single-crystal
interferometer is at rest with respect to the laboratory, whereas the
neutrons are subject to the gravitational potential. Bonse and
Wroblewski (BW) \cite{BW} compared this with the effect of {\it
  acceleration} relative to the laboratory frame by letting the
interferometer oscillate horizontally. With these experiments of BW
and COW the effect of local acceleration and local gravity on matter
waves has been shown to be equivalent. Later, with atomic beam
interferometry, the accuracy of these type of results were appreciably
improved.

It is strange, but in most textbooks on gravitation---and in most
philosophical discussions on gravity---these successful experiments on
the behavior of Dirac fields under acceleration (BW) and in a
gravitational field (COW) are simply not mentioned. Most textbook
authors and philosophers rather restrict themselves to Einstein's 1916
discussion and to experiments related therewith. In writing a
textbook on gravitation, is it indecent to refer to experiments that
have a certain quantum flavor? Is it appropriate to be silent about
experiments that provide new insight into the structure of the
gravitational field?

The neutrons in the COW and BW experiments have spin $\frac 12$, they
are fermions. At the energies prevalent in the COW and the BW
experiments, the neutron (including its spin) can be supposed to be
elementary, its composition out of three quarks can be
neglected. Accordingly, if the neutron is force-free, it can be
described by a Dirac spinor $\Psi(x)$ obeying the free Dirac
equation\footnote{Here $\hbar=1, c=1$, the imaginary unit is denoted
  by $i$, the Dirac gamma matrices by $\gamma^k$, and the mass of the
  neutron by $m$. If an electromagnetic field is present, the Dirac
  equation has to be coupled minimally to it and a Pauli-term added
  that takes into account the non-standard magnetic moment of the
  neutron.} $ (i\gamma^k\partial_k-m)\Psi(x)= 0$. Thus, the neutron
obeys approximately a classical one-particle equation, namely the
Dirac or, in the non-relativistic limit, the Pauli-Schr\"odinger
equation and, if the spin can be neglected, the Schr\"odinger
equation. That this evaluation is correct has been borne out by
experiments of the COW and BW type \cite{RauchW}: the neutrons of the
COW and the BW experiments obey a Schr\"odinger equation including a
Newtonian gravitational potential energy or a corresponding
acceleration term, respectively.

\begin{figure}
\begin{center}
\includegraphics[width=10truecm]{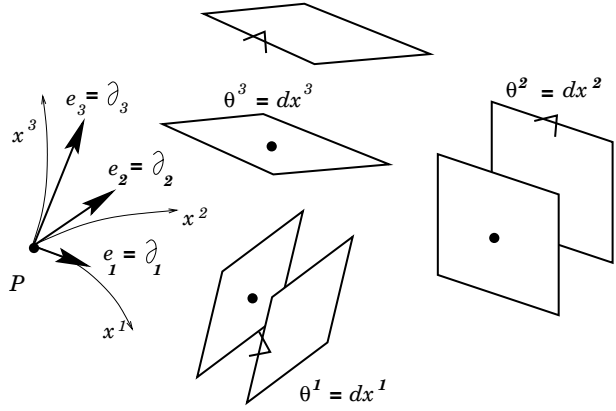}
\caption{Natural frame $e_b=\d_b ^j\partial_j$ and natural coframe
  $\vt^a=\d^a_i dx^i$ at a point $P$ of a three-dimensional manifold
  ($a,b=1,2,3$). The coordinates of $P$ are denoted by $x^i$,
  $i=1,2,3$, whereas $\d_a^b$ is the the Kronecker symbol. The coframe
  $\vt^a$ is supposed to be also at the same point $P$, but the three
  one-forms $\vt^a$ are shifted for better visibility in 3 different
  directions. Note that $\vartheta^1(e_1)=1,\, \vartheta^1(e_2)=0$,
  etc., that is, $\vartheta^a$ is dual to $e_b$ according to
  $e_b\lrcorner \vt^a \equiv\vartheta^a(e_b)=\delta^a_b$; for the
  figure, see \cite{Birkbook}.}
\end{center}
\end{figure}

The basic difference between the mass point and the Dirac field is
that the latter requires an {\it orthonormal} reference frame for its
description. A Dirac spinor is a half-integer representation of the
[covering group $SL(2,C)$ of the] Lorentz group $SO(1,3)$, that is, it
is intrinsically tied to the Lorentz group. In Minkowski space it is
simple to introduce an orthonormal frame. On starts with Cartesian
coordinates and takes the tangent vectors of the coordinate lines as
``natural'' frame $e_\b=\d_\b ^j\partial_j$, compare Figure~1. If one
translates and Lorentz rotates such a frame, one can find an arbitrary
frame $e_\b=e^j{}_\b\partial_j$ that, in general, cannot any longer
be derived {}from coordinate lines. Before we discuss this {}from a more
general point of view, let us first make a general remark:

Pitts \cite{Pitts:2011jv} argues, using work of Ogievetsky \&
Polubarinov of the 1960s, that one doesn't require orthonormal frames
for introducing spinors in curved spacetime and that coordinate
systems are sufficient. Frames are very useful for Fermi-Walker
transport and for gravitomagnetism already in GR. For the gauge theory
of gravity, frames were used by Sciama and Kibble, see \cite{Reader},
and we can hardly see a benefit for kicking them out. The price one
has to pay for the removal of frames is to go to nonlinear group
representations and to other complications. We do not know whether
this prevention of frames is really conclusive and leave the answer to
this question to the future.

\section{Some geometric machinery: coframe and connection}

Suppose that spacetime is a four-dimen\-sional continuum in which we
can distinguish one time and three space dimensions. At each point
$P$, we can span the local cotangent space by means of four {\it
  linearly independent} covectors, the {\it coframe\/}
$\vta^\a=e_i{}^\a dx^i$. Here $\a,\beta ,\dots=0,1,2,3$ are frame and
$i,j,\dots =0,1,2,3$ coordinate indices. In general, the object of
anholonomity two-form does not vanish,
\begin{equation}\label{anholonomity}
C^\a:=d\vt^\a={\scriptstyle\frac 12}  C_{ij}{}^\a dx^i\wedge
dx^j \ne 0\,,\qquad\text{with}\qquad
C_{ij}{}^\a=2\partial_{[i}e_{j]}{}^\a\,,
\end{equation}
see \cite{SchoutenPhys}. This specification of spacetime is the bare
minimum that one needs for applications to classical physics.

As soon as we have a coframe $\vt^\a$, we can also define its dual,
the frame composed of four likewise linearly independent vectors
$e_\a=e^i{}_\a\partial_i$ by the duality relation $e_{\b} \lrcorner
\vt^a=\vt^\a(e_\b)=\d^\a_\b$. Geometrically speaking, frame and
coframe are equivalent as reference frames for physical quantities. For
physical reasons, the coframe turns out to be the translational gauge
type potential and thus does fit more smoothly into a gauge formalism.

Having now a reference coframe $\vt^\a$, we want to do physics in such
a spacetime. We need a tool to express, for instance, that a certain
field is constant. If the field is a scalar $\phi$, there is no
problem, the gradient $d\phi=(\partial_i\phi) dx^i$, if equated to zero, will do
the job.  However, if the field is a vector or, more generally, a
spinor or an arbitrary tensor field $\psi$, we need a law that
specifies the parallel transfer of $\psi$ {}from one point $P$ to a
neighboring point $P'$. Let us see how Einstein in 1955 looked in
retrospect at the development of GR \cite{Einstein:1955}:
\begin{quotation}
  ...the essential achievement of general relativity, namely to
  overcome `rigid' space (ie the inertial frame), is {\it only
    indirectly} connected with the introduction of a Riemannian
  metric. The directly relevant conceptual element is the
  `displacement field' ($\Gamma^l_{ik}$), which $\,$expresses the$\,$
  infinitesimal displacement of vectors. It is this which replaces the
  parallelism of spatially arbitrarily separated vectors fixed by the
  inertial frame (ie the equality of corresponding components) by an
  infinitesimal operation. This makes it possible to construct tensors
  by differentiation and hence to dispense with the introduction of
  `rigid' space (the inertial frame). In the face of this, it seems to
  be of secondary importance in some sense that some particular
  $\Gamma$ field can be deduced {}from a Riemannian
  metric...\footnote{When I showed this quotation during my seminar,
    E.~Scholz (Wuppertal) immediately remarked that the fact of the
    importance of the connection as guiding field was already clear to
    Weyl in 1918, or at least in the 1920s. And D.~Rowe (Mainz) added
    that also Einstein was aware of the importance of the concept of a
    connection since at least the late 1920s. Both remarks are
    certainly true. However, there is a subtle difference: Weyl
    referred to a symmetric connection since he was concerned with
    coordinates and not with frames. When, in 1929, he introduced
    frames \cite{Weyl:1929}, Weyl's connection still remained
    symmetric, and only in 1950 he considered also {\it a}symmetric
    connections in the context of gravity \cite{Weyl:1950}. In
    contrast, Einstein was concerned with {\it a}symmetric connections
    at least since 1925, when he formulated a unified theory of
    gravity and electricity and introduced what is nowadays called
    incorrectly the Palatini variational principle
    \cite{Einstein:1925}.}
\end{quotation}
Einstein's `displacement field' can be implemented by means of a {\it
  linear connection\/} $\Gamma_\a{}^\beta = \Gamma_{i\a}{}^\beta dx^i$
(``affinity''). The one-form field $\Gamma_\a{}^\beta (x)$, with its 64
independent components, has to be prescribed before the parallel
transport of a spinor or a tensor field $\psi$ can be performed and,
associated with it, a covariant derivative be defined (whose vanishing
would imply that the field is constant). The linear connection
$\Gamma_\a{}^\beta (x)$, shortly after the advent of general
relativity, was recognized as a fundamental ingredient of spacetime
physics, for more details see \cite{Reader}, for instance. The law of
parallel transport embodies the {\it inertial properties\/} of matter.

The connection ${\Gamma_{\a}{}^\b }$ represents $4\times 4$ potentials
of the four-dimensional group of general linear transformations
$GL(4,R)$. Very similar to the Yang--Mills potential of the $SU(3)$,
for example.

Coframe and connection $\vt^a,\Gamma_\a{}^\b$---still the metric is
not involved---provide a good arsenal for further geometrical
battles. Having a connection, we can covariantly differentiate. We
define straightforwardly the ``field strengths'' torsion ${T^\a}$ and
curvature ${R_\a{}^\b}$ as
\begin{eqnarray}\label{tor}{ T^\a}&:=&d \,{\vta^\a}+
  {\Gamma_\b{}^\a}\wedge{\vta^\b}
  \hspace{16pt}=\;{\scriptstyle\frac
  12} T_{ij}{}^\a dx^i\wedge dx^j\,,\\ \label{curv}
  { R_\a{}^\b}&:=&d\,{\Gamma_\a{}^\b}
  -{\Gamma_\a{}^\g}\wedge{\Gamma_\g{}^\b}
  \hspace{5pt}=\;{\scriptstyle\frac 12} R_{ij\a}{}^{\b}dx^i\wedge dx^j\,.
\end{eqnarray}
One recognizes that ${T^\a}$ and ${R_\a{}^\b}$ are the gauge field
strengths of the affine group $A(4,R)=T(4)\!\rtimes\!{GL(4,R)}$.

Let us look at the torsion in components. {}From (\ref{tor}) we find
\begin{equation}
  T_{ij}{}^\a=2\partial_{[i}e_{j]}{}^\a+2\Gamma_{[i|\b}{}^\a e_{|j]}{}^\b=
  C_{ij}{}^\a+2\Gamma_{[ij]}{}^\a\,.
\end{equation} 
In a holonomic (coordinate) frame, $C_{ij}{}^\a=0$. Thus, $T_{ij}{}^\a
\stareq 2\Gamma_{[ij]}{}^\a$; incidentally, a `star equal' $\stareq$
is used, see \cite{SchoutenPhys}, if a formula is only valid for a restricted
class of frames or coordinates. In such a frame---and only in a
holonomic one---the vanishing of the torsion translates into the {\it
  symmetry of the connection.} It is now obvious why this symmetry is
called a ``bastard symmetry'': in
$\Gamma_{[ij]}{}^\a=\Gamma_{[i|\b}{}^\a e_{|j]}{}^\b$, the index `i'
originates {}from the one-form character of the connection, whereas
the index `j' is related to the Lie-algebra index `$\b$'. Only in a
holonomic frame the symmetry of a connection looks natural. In an
anholonomic frame, here $C_{ij}{}^\a\ne 0$, it is nothing trivial. It
is a fundamental assumption that has to be justified similar as the
vanishing of the curvature.

A space with $T^\a\ne 0,\,R_\a{}^\b\ne 0$, we call an {\it affine}
space. If $T^\a=0$, we have a {\it symmetric} affine space, if
$R_\a{}^\b=0$, we have a {\it teleparallel} affine space (or of a
space with teleparallelism). Should we require $T^\a=0$ and
$R_\a{}^\b=0$, we have a symmetric flat affine space.

We followed here the lead of Schr\"odinger \cite{SchrodingerSTS} and
introduced first the connection before we will turn to the metric.

\section{More geometry: metric and orthonormal coframe}

However, our experience in Minkowski space tells us that there must be
more structure on the spacetime mani\-fold than the symmetric flat
affine space possesses. Locally at least, we are able to measure time
and space intervals and angles. A pseudo-Riemannian (or Lorentzian)
metric\footnote{Nowadays there exists a definite hint that the
  conformally invariant part of the metric, the light cone, is
  electromagnetic in origin (see \cite{Birkbook,Hehl:2005hu}), that
  is, it can be derived {}from premetric electrodynamics together with
  a linear constitutive law for the empty spacetime (vacuum). Hence
  the metric, or at least its conformally invariant part, doesn't
  appear as a fundamental structure, it rather emerges in an
  electromagnetic context.}  $g_{ij} =g_{ji}$ is sufficient for
accommodating these measurement procedures.  If $g_{\a\beta }$ denotes
the components of the metric with respect to the coframe, we have
$g_{ij}=e_i{}^\a e_j{}^\beta g_{\a\beta }$ and
${\mathbf{g}}={g_{\a\b}}\,{\vta^\a}\!\otimes{\vta^\b}$. In an
orthonormal coframe we recover
\begin{equation} \label{metric} g_{\a\b}\ \stareq\ o_{\a\b}\ :=
  \left(\begin{array}{crrr} 1 & 0 & 0 & 0\\ 0 &-1 & 0 & 0\\ 0 & 0 &-1
      & 0\\ 0 & 0 & 0 &-1 \end{array}\right).
\end{equation}

Now, in analogy to the procedures in equations (\ref{tor}) and
(\ref{curv}), we can derive the field strength, the {\it nonmetricity}
one-form, corresponding to the potential $g_{\a\b}$, by
differentiation:
\begin{equation}\label{nonm}
  Q_{\a\b}:=-Dg_{\a\b}=-dg_{\a\b}+\Gamma_\a{}^\g g_{\g\b}+\Gamma_\b{}^\g
  g_{\a\g}=Q_{i\a\b}dx^i\,.
\end{equation}
Accordingly, the coframe $\vartheta^\a(x)$, the linear connection
$\Gamma_\a{}^\beta(x) $, and the metric $g_{\a\beta }(x)$ control the
geometry of spacetime. The metric determines the distances and angles,
the coframe serves as translational gauge potential, whereas the
connection provides the guidance field for matter reflecting its
inertial properties and it is the $GL(4,R)$ gauge potential. The space
equipped with these $10+16+64$ potentials
$(g_{\a\b},\vt^\a,\Gamma_\a{}^\b)$ we call a {\it metric-affine}
space, the corresponding field strength are the $40+24+96$ fields
$(Q_{\a\b},T^\a,R_\a{}^\b)$, for reviews and the corresponding
formalism, see \cite{Reader,Erice95,PRs}.

In a metric-affine space, we can lower the second index of the
connection according to $\Gamma_{\a\b}:=\Gamma_\a{}^\g g_{\g\b}$. Then
we can compare it with the Riemann (Levi-Civita) connection
$\widetilde{\Gamma}_{\a\b}$. After some algebra, see
\cite{SchoutenPhys}, we find in terms of components:
\begin{equation}\label{magConn}
  {\Gamma}_{\a\b\g}=\widetilde{\Gamma}_{\a\b\g} +{\scriptstyle\frac
    12}(T_{\a\b\g} -T_{\b\g\a}+T_{\g\a\b})
  +{\scriptstyle\frac 12}(Q_{\a\b\g} +Q_{\b\g\a} -Q_{\g\a\b})\,.
\end{equation}
It should be stressed that this decompositions are useful if a direct
comparison is made with the Riemannian piece
$\widetilde{\Gamma}$. However, in the variational formalism of a gauge
theory of gravity, besides $g_{\a\b}$ and $\vt^\a$, the connection
$\Gamma_\a{}^{\beta}$ is considered as {\it independent}
variable. Then such a decomposition is unwarranted under those
circumstances.

Can we give a satisfactory justification for the emergence of
three different gravitational gauge potentials? We take the Minkowski
space of SR as basis for our considerations. It is a
fact of life that the geometry of a Minkowski (or a Euclidean) space
consists of an interplay between properties that relate to parallel
displacement and those that relate to distance and angle
measurements. In Minkowski space this duality between affine
(inertial) and metric properties is solved in that the affine
properties are exclusively expressed in terms of metric properties:
the metric properties dominate the affine ones. 

If we ``liberate'' the affine properties, we are immediately led, in
four dimensions, to the affine group $A(4,R)=T(4)\!\rtimes\! GL(4,R)$
and, gauging it, to the coframe $\vt^\a$ and the linear connection
$\Gamma_\a{}^\b$ as gauge potentials. The metric properties, expressed
by the metric $g_{ij}$, are then left behind. 

Since macroscopic gravity in GR is so successfully described by means
of the metric $g_{ij}$ as (Einstein's) gravitational potential, it
suggests itself to add the metric---in its anholonomic form
$g_{\a\b}$---as third member to the gravitational potentials. There
are two procedures possible: We pick, instead of an arbitrary coframe,
an {\it orthonormal} one, which is constructed with the help of the
metric; in this way the metric is absorbed and, besides this
orthonormal coframe, only the connection remains as variable. However,
since this restricts the freedom of choosing also non-orthonormal
coframes, we take all three potentials as independent variables. The
Lagrangian formalism of the corresponding field theory will then
provide the relation between the coframe and the metric, and it will
turn out that there is, indeed, a close link between both variables,
see \cite{PRs}. At the same time---and this is a real progress in
understanding---we find that the metric energy-momentum current of
matter $\mathfrak{t}^{\a\b}$ couples to the metric and the canonical
one $\mathfrak{T}_\a$ couples to the coframe. Their interdependence is
beautifully displayed in the three-potentials' approach.

\begin{figure}
\begin{center}
\includegraphics[width=9truecm]{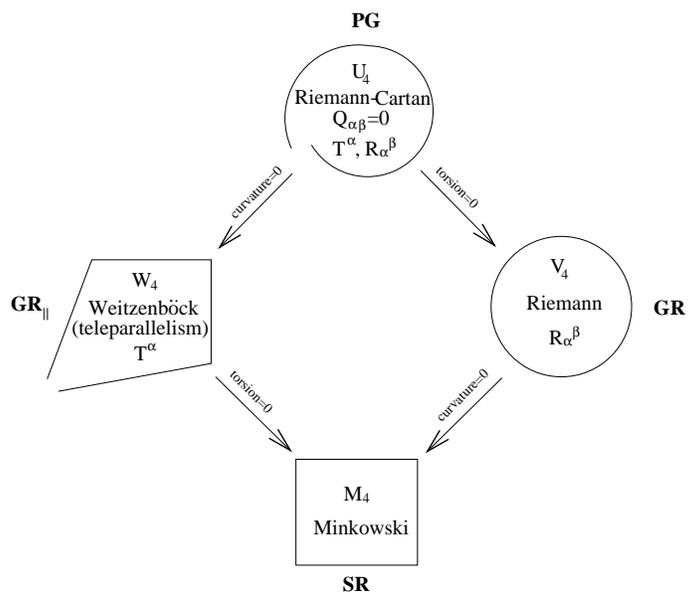}
\end{center}
\caption{The Riemann--Cartan space (or $U_4$), a metric-affine space
  with vanishing nonmetricity, is the arena for the Poincar\'e gauge
  theory (PG). It can either become a {\it Weitzenb\"ock} space $W_4$,
  if its curvature vanishes, or a {\it Riemann} space $V_4$, if the
  torsion happens to vanish. GR acts in a $V_4$, teleparallelism
  theories of gravity in a $W_4$.}
\end{figure}

In a metric-affine space, as shown by Hartley \cite{Hartley}, {\it
  normal frames} can be found: {\it locally} it is possible to find
suitable coordinates and suitable frames such that
\begin{equation}\label{normal}
  (\vt^\a,\Gamma_\a{}^\b)\stareq (\d^\a_i dx^i, 0)\,.
\end{equation}
This is the new type of Einstein elevator. In GR, the Einstein
elevator was described by a holonomic reference frame $\vt^\a$ with
$C^\a=0$. Then, in the Riemann spacetime of GR, one could introduce
Riemannian normal coordinates. Here, in the gauge theoretical
approach, the constraint of holonomicity is dropped and this new
degree of freedom, which expresses itself in a rotational
acceleration, admits to introduce normal frames. The equivalence
principle can then be applied in this new context. For new
developments of this notion, see Nester \cite{JimNormalFrames} and
Giglio \& Rodrigues \cite{Giglio:2011gk}.

As soon as we require in a metric-affine space integrability of length
and angle measurements, we have to postulate\footnote{If one wants to
  keep the angles integrable, but not the length, one can postulate
  only the vanishing of the tracefree part of the nonmetricity,
  $Q_{\a\b}-{\scriptstyle\frac 14} g_{\a\b}Q_\g{}^\g=0$. This results
  in a Weyl--Cartan space with non-vanishing Weyl covector
  ${\scriptstyle\frac 14}Q_\g{}^\g$, see the contribution of Scholz
  \cite{Scholz}; however, in this approach also the torsion is put to
  zero.}  $Q_{\a\b}=0$. Then we arrive at a {\it Riemann--Cartan} space
(RC-space), which was mentioned in the context of the gauging of the
Poincar\'e group in Sec.~3. In such a space, if we choose {\it
  orthonormal} frames, the connection becomes antisymmetric; then we
have the $6+ 24$ potentials $(\vt^a,\Gamma^{\a\b}= -\Gamma^{\b\a})$ as
gravitational variables. Again normal frames like in (\ref{normal})
can be found, 
\begin{equation}\label{KibbleLab}
(\vt^\a,\Gamma^{\a\b})\stareq (\d^\a_i dx^i, 0)\,,
\end{equation}
as has been first shown by von der Heyde \cite{vdHnormal}. This
geometrical fact shows clearly that the Lorentz connection
$\Gamma^{\a\b}$ is, besides the orthonormal frame $\vt^\a$, the
appropriate gauge field variable. After this geometrical detour we are
back to where we started {}from. In Figure~2 different subcases of a
RC-space are displayed.

\section{Dirac field $\Psi(x)$ in Minkowski space in a non-inertial
  reference frame}

After this rather long geometrical interlude, we come back to physics
and consider again the Dirac field $\Psi(x)$. A mass point in an
inertial frame moves according to equation (\ref{eqmo1}), one in an
accelerated frame, according to equation (\ref{eqmo2}). The inertial
forces are represented by the Christoffels in equation
(\ref{Christ}). Let us execute the analogous process for the Dirac
electron. Since the Dirac electron is referred to an orthonormal
(co)frame, we have to study its behavior under translational and
rotational accelerations, see \cite{HehlNi}.

In Minkowski space in Cartesian coordinates, we have the force-free
Dirac equation as analog of equation (2),
\begin{equation}\label{x}
(i\gamma^i\partial_i-m)\Psi\stareq 0\,,
\end{equation}
and in a non-inertial frame in flat Minkowski space we find 
\begin{equation}\label{y}
  \left[i\gamma^\alpha e^i{}_\alpha(\partial_i+{\frac{i}{4}}\,
\sigma_{\beta\gamma}\widetilde{\Gamma}_i{}^{\beta\gamma})
-m\right]\!\Psi=0\,,\qquad \sigma_{\b\g}:=i\gamma_{[\b}\gamma_{\g ]}  \,.
\end{equation}
These two equations correspond to the Einsteinian equations
(\ref{eqmo1}) and (\ref{eqmo2}). Namely, in the non-relativistic
WKB-approximation, when the spin can be neglected, equation (\ref{x})
becomes (\ref{eqmo1}). You may wonder whether this is true since
(\ref{x}), in contrast to (\ref{eqmo1}), is mass dependent. For this
reason some people argued that this violates the equivalence principle
in the sense that the motion of a force-free particle (field) must be
independent of $m$. However, what they overlooked is that also in
classical mechanics the Hamilton-Jacobi equation for a force-free
particle {\it is} mass dependent---and the classical non-relativistic
analog of the Dirac equation is the Hamilton-Jacobi
equation. Accordingly, all is fine and in the desired approximation
the mass will drop out.

The new potentials, emerging in a non-inertial frame, are
$(e^i{}_\a,\widetilde{\Gamma}_i{}^{\b\g})$. The latter one, in
Minkowski space, can be expressed in terms of derivatives of the
former: $\widetilde{\Gamma}_i{}^{\b\g}= \widetilde{\Gamma}_i{}^{\b\g}
(\partial_j e^k{}_\d)$. However, we will not substitute
$\widetilde{\Gamma}_i{}^{\b\g}$ in terms of the the frame since we
will relax the constraint $T_{ij}{}^\a=0$ subsequently.

This is what we will do now. Einstein relaxed the constraint
$\widetilde{R}_{ijk}{}^\ell=0$, since that is all he found for a point
particle, we relax the constraints $T_{ij}{}^\a=0$ and
$\widetilde{R}_{ij}{}^{\a\b}=0$, since a Dirac field has a more
involved structure as displayed in particular in a non-inertial
frame. This relaxation of both constraints leads directly to a {\it
  Riemann--Cartan spacetime} as the arena appropriate for a Poincar\'e
gauge theory (PG).

Why couldn't we do by only relaxing the curvature constraint,
$\widetilde{R}_{ij}{}^{\a\b}\ne 0$, but keeping the torsion
constraint, $T_{ij}{}^\a=0$? Well, this is possible. However, it is
not in the sense of local field theory. Why should we keep the
non-local constraint\footnote {Explicitly, this constraint reads
  $T_{ij}{}^\a=2(\partial_{[i} e_{j]}{}^\a+\Gamma_{[i|\b|}{}^\a
  e_{j]}{}^\b)=0$. These are $6\times 4=24$ PDEs for the coframe
  components $e_i{}^\a$. For their solution, we not only have to know
  the local values of $e_i{}^\a$, but also their values in the
  infinitesimal neighborhood. In this sense, the constraint is
  non-local and contrived, see \cite{vdHnormal} for a more detailed
  discussion.} $T_{ij}{}^\a=0$, which corresponds to 24 partial
differential equations of first order, when we know that its
relaxation does away with these PDEs and still allows locally to get
rid of gravity according to (\ref{KibbleLab})?

Whereas Einstein discussed the equivalence principle on the level of
the equations of motion, in gauge theories, because of the application
of the Noether theorem for rigid and local symmetries, the discussion
takes place on the level of Lagrangians. If we multiply
$D_\a\Psi=(\partial_\a +\frac i4 \sigma_{\b\g}\Gamma_\a{}^{\b\g})\Psi$
{}from the left by $i\overline{\Psi}\gamma^\a$, average with its
Hermitian conjugate, and add a mass term, we find the (real) Dirac
Lagrangian density:\footnote{$e:=\det
  e_i{}^\a,\> \partial_\a=e^i{}_\a\partial_i,\> D_\a=e^i{}_\a D_i$.}
\begin{eqnarray}\label{DiracLagrangian}
 \frac{\mathfrak L}{e}&=&\frac i2 e^i{}_\a\left[\overline{\Psi}\gamma^\a
    \left(\partial_i+\frac
      i4\sigma_{\b\g}\Gamma_i{}^{\b\g}\right)\Psi \right]\>
  +\>\text{herm.\ conj.}+m\overline{\Psi}\Psi
  \nonumber\\ &=&\frac
  i2(\overline{\Psi}\gamma^\a D_\a\Psi
  -\Psi\gamma^\a\overline{D_\a\Psi})+m\overline{\Psi}\Psi\,.\label{DiracLagr}
\end{eqnarray}
The action is $W=\int d^4x \mathfrak{L}$. The Lagrangian in an
inertial frame in Cartesian coordinates can be read off by making the
substitutions
$e^i{}_\a\rightarrow\d^i_\a,\,\Gamma_i{}^{\b\g}\rightarrow 0$.

\section{Some results of the Lagrange--Noether formalism}

To identify the currents that couple to the gravitational potentials
$(e_i{}^\a,\Gamma_i{}^{\a\b})$, some formalism is necessary that may
disturb the philosophically minded reader. We try to simplify these
considerations and will, instead of working in a RC-spacetime (for a
rigorous treatment see \cite{PRs}), restrict ourselves to the
Minkowski space in Cartesian coordinates. 

The action $W$ is invariant under 4 rigid spacetime {\it translations}
of 6 rigid Lorentz {\it rotations} (3 boosts plus 3 spatial
rotations). As a consequence, we have (see Corson \cite{Corson})
energy-momentum and angular momentum conservation,
\begin{equation}\label{cons}
\partial_k\mathfrak{ T}_i{}^k\stareq 0\,,\qquad\partial_k {\mathfrak
  J}_{ij}{}^k\stareq 0\,,
\end{equation}
with the {\it canonical energy-momentum} tensor density
\begin{equation}\label{canT}
\mathfrak{T}_i{}^k:\stareq\d_i^k\mathfrak{L}-\frac{\partial\mathfrak{L}}
{\partial\partial_k\Psi}\partial_i\Psi\,,
\end{equation}
the total canonical angular momentum tensor density, consisting of an
intrinsic and an orbital part,
\begin{equation}\label{canJ}
  \mathfrak{J}_{ij}{}^k:\stareq  \mathfrak{S}_{ij}{}^k+x_i\mathfrak{T}_j{}^k-
x_j\mathfrak{T}_i{}^k=- \mathfrak{J}_{ji}{}^k,
\end{equation}
and the {\it canonical spin} angular momentum tensor density
($l_{ij}$=Lorentz generators)
\begin{equation}\label{canS}
 \mathfrak{S}_{ij}{}^k:\stareq\frac{\partial\mathfrak{L}}
{\partial\partial_k\Psi}l_{ij}\Psi=- \mathfrak{ S}_{ji}{}^k\,.
\end{equation}

{}From this straightforward consideration in Minkowski space alone, we
recognize that the canonical energy-momentum $\mathfrak{T}_i{}^k$ and
the canonical angular momentum $\mathfrak{J}_{ij}{}^k$ are the
translational and the Lorentz currents of matter. Only the intrinsic
spin part $ \mathfrak{S}_{ij}{}^k$ of the angular momentum is a
tensor; the orbital part is only a tensor under Cartesian coordinate
transformations. For the Dirac field we find ($\mathfrak{T}_{\,\a}{}^k
=e^i{}_\a\mathfrak{T}_{\,i}{}^k, \text{etc.}$):
\begin{eqnarray}\label{DiracEM}
  \mathfrak{T}_\a{}^i&\stareq&\frac{i}{2}\left(\overline{\Psi}
    \gamma^i\partial_\a\Psi-
    \Psi\gamma^i\partial_\a\overline{\Psi} \right)\,,\\ \label{DiracSpin}
  \mathfrak{S}_{\a\b}{}^i&\stareq&\frac
  18\overline{\Psi}\left(\sigma_{\a\b}\g^i+\g^i\sigma_{\a\b}\right)\Psi\,.
\end{eqnarray}
The spin is totally antisymmetric: after some algebra, we can
put (\ref{DiracSpin}) into the form
\begin{equation}\label{skewspin}
  \mathfrak{S}_{\a\b\g}=\frac 14
  \epsilon_{\a\b\g\d}\overline{\Psi}\g^\d\g_5\Psi\,,
\end{equation}
with $\g_5:=-\frac{i}{4!} \epsilon_{\a\b\g\d} \g^\a \g^\b \g^\g \g^\d$.

We compare (\ref{DiracEM}) and (\ref{DiracSpin}) with the Lagrangian
(\ref{DiracLagr}) and consider small deviations {}from the inertial
case, that is, $e^i{}_\a=\d^i{}_\a+\epsilon^i{}_\a$, with
$\epsilon^i{}_\a\ll 1$, then we find after some algebra and some
rearrangements to linear order in $\epsilon^i{}_\a$,
\begin{eqnarray} \label{lin}
{\mathfrak L}&\sim&e_i{}^\a\mathfrak{T}_\a{}^i
  +  \Gamma_i{}^{\a\b}\mathfrak{S}_{\a\b}{}^i
  -m\overline{\Psi}\Psi\,.
\end{eqnarray}
There is some resemblance to the structure in (\ref{eqmo3}) even
though we work here on a Lagrangian level. This coupling of geometry to
matter displayed in (\ref{lin}) suggests the following representation
of the canonical currents:
\begin{equation}\label{dynTS}
  \mathfrak{T}_\a{}^i=\frac{\d\mathfrak{L}}{\d e_i{}^\a}\,,\qquad 
  \mathfrak{S}_{\a\b}{}^i=\frac{\d\mathfrak{L}}{\d \Gamma_i{}^{\a\b}}\,.
\end{equation}

Of course, this was a heuristic consideration, but with the full
Lagrange-Noether machinery acting in RC-spacetime, it can be made
rigorous \cite{PRs}: The canonical currents $\mathfrak{T}_\a{}^i,\,
\mathfrak{S}_{\a\b}{}^i$, defined via the Noether theorem according to
(\ref{canT}) and (\ref{canS}), can be shown to be equal to the
``dynamical'' currents that couple to the gravitational potentials
according to (\ref{dynTS}). These currents should also play a decisive
role in quark and gluon physics, see \cite{Hehl:2014eja}.\bigskip

\noindent{\sl A short summary of the formalism in this section}\medskip

For those of you who were lost in this formalism, a short bird eye's
view on the results: In order to compactify our notation, we change to
exterior calculus. We introduce the matrix-valued one-form
$\gamma:=\gamma_\a \vt^\a$ and the Hodge star operator $^\star$. Then
the Dirac equation in an arbitrary orthonormal frame in a RC-space can
be rewritten as
\begin{eqnarray}
i{}^\ast\gamma\wedge D\Psi + {}^\ast  m\,\Psi =0\,,\label{DiracEx}
\end{eqnarray}
with the covariant exterior derivative $D\Psi:=(d
+\frac{i}{4}\sigma_{\a\b}\Gamma^{\a\b})\Psi$.  Let us then formulate
Lagrange four-form of the Dirac field,
\begin{equation}
L=L(\vartheta^\alpha,\Psi,D\Psi)=
\frac{i}{2}\left(\overline{\Psi}\,{}^\ast\gamma\wedge D\Psi
+\overline{D\Psi}\wedge{}^\ast\gamma\,\Psi\right)+{}^\ast m\,
\overline{\Psi}\Psi\,,
\label{DiracLagrangian*}
\end{equation}
which is minimally coupled to the RC-spacetime via the gauge
potentials $\vt^\a$ [contained in $\g=\g_\a\vt^\a$] and
$\Gamma^{\a\b}=-\Gamma^{\b\a}$ [contained in $D$]. Note that only the
potentials themselves enter the Lagrangian, but {not} their
derivatives. Thus, the Lagrangian (\ref{DiracLagrangian*}), formulated
in a RC-spacetime, because of (\ref{KibbleLab}), looks {\it locally}
special-relativistic. This attests to the validity of the relaxation
process discussed above. The currents, as we saw above in
(\ref{dynTS}), are then defined as follows:
\begin{equation}\label{curr}
\mathfrak{T}_\a=\frac{\d L}{\d \vt^\a}\,,\qquad 
\mathfrak{S}_{\a\b}=\frac{\d L}{\d \Gamma^{\a\b}}\,.
\end{equation}
These innocently looking equations (\ref{DiracLagrangian*}) and
(\ref{curr}), all living in a RC-spacetime, are the net outcome of our
considerations so far. 

It was then Sciama \cite{Sciama} and Kibble \cite{Kibble:1961} in the
early 1960s who added the Hilbert--Einstein type Lagrangian of the
RC-spacetime to (\ref{DiracLagrangian}) and formulated the
corresponding simplest field equations of the gauge theory of gravity;
for a historical view see O'Raifeartaigh \cite{Lochlainn} and the
reprint volume \cite{Reader}, for a modern representation Blagojevi\'c
\cite{Milutinbook} and Ryder \cite{Ryder}.

 \section{Field equations of Sciama and Kibble} 

 The Ricci tensor in a RC-spacetime is defined according to
 $\text{Ric}_i{}^\a:=e^j{}_\b R_{ji}{}^{\a\b}$. A corresponding scalar
 density $ee^i{}_\a \text{Ric}_i{}^\a$ is the simplest nontrivial
 gravitational Lagrangian. The total action is ($\Lambda$ =
 cosmological constant)
\begin{equation}\label{action}
  W_{\text{tot}}=\int d^4x\left[{\frac{1}{2\kappa}} e(e^i{}_\a 
    \text{Ric}_i{}^\a -2\Lambda)+\mathfrak{L}(e_k{}^\g,\Psi,D\Psi)\right]\,,
\end{equation}
with Einstein's gravitational constant $\kappa$. Variation with
respect to $e_i{}^\a$ and $\Gamma_i{}^{\a\b}$ yields the gravitational
field equations of Sciama \cite{Sciama} and Kibble \cite{Kibble:1961}:
\begin{eqnarray}\label{firsta}
  {\rm Ric}_\a{}^i-{\scriptstyle\frac{1}{2}}e^i{}_\a\, {\rm Ric}_\g{}^\g 
+\Lambda e^i{}_\a&=& {\frac{\kappa}{e}} \,\frak{T}_\a{}^i\,,\\
  \label{seconda}
  \text{Tor}_{\a\b}{}^i- e^i{}_\a \text{Tor}_{\b\g}{}^\g+ e^i{}_\b 
  \text{Tor}_{\a\g}{}^\g
  &=& {\frac{\kappa}{e}}\, \frak{S}_{\a\b}{}^i\,.
\end{eqnarray}
We made here the torsion a bit more visible. Please note that Ric and
$\mathfrak{T}$ have both 16 independent components, whereas Tor and
$\mathfrak{S}$ have both 24 independent components. These field
equations are just linear algebraic equations between Ric and Tor on
the geometrical side and $\mathfrak{T}$ and $\mathfrak{S}$ on the
matter side, respectively. The Dirac case is particularly simple,
there (\ref{seconda}) collapses to just 4 equations.

The first equation can be easily recognized as an Einstein type field
equation. However, the Ricci tensor is here asymmetric as well as the
canonical energy-momentum tensor of matter. The second equation
relates the torsion linearly to the spin of matter. If we consider
matter {\it without} spin, the torsion vanishes and the first field
equation reduces just to the Einstein field equation of GR, for a
review see \cite{RMP}.

In exterior calculus, these field equations, given first in this form
by Trautman, see \cite{Trautman}, look even a bit more
transparent:\footnote{Here we have: Hodge star $^\star$,
  $\eta_\a={}^\star\vt_a,\,\eta_{\a\beta}={}^\star(\vt_a\wedge\vt_\beta),\,
  \eta_{\a\beta\g}={}^\star(\vt_a\wedge\vt_\beta\wedge\vt_\g)$. Moreover,
  $\mathfrak{T}_\a={\scriptstyle\frac{1}{e}}\mathfrak{T}_\a{}^\g
  \eta_\g$ and $\mathfrak{S}_{\a\b}={\scriptstyle
    \frac{1}{e}}\mathfrak{S}_{\a\b}{}^\g \eta_\g$.}
\begin{eqnarray}\label{firstb}
{\scriptstyle\frac{1}{2}}\eta_{\a\beta\g}\wedge R^{\beta\g}-\Lambda \eta_\a
&=& \kappa \,\frak{T}_\a\,,\\
{\scriptstyle\frac{1}{2}}\eta_{\a\beta\g}\wedge T^{\g}
&=&\kappa\,
\frak{S}_{\a\beta}\,.
\label{secondb}
\end{eqnarray}
The two equations (\ref{firsta}),(\ref{seconda}) or
(\ref{firstb}),(\ref{secondb}) are the field equations of the
Einstein--Cartan(--Sciama--Kibble) theory of gravity or, in short, of
the {\it Einstein--Cartan theory} (EC). This is a special case of a
Poincar\'e gauge theory, namely that which has the curvature scalar of
the RC-spacetime as gravitational Lagrangian. EC is a viable
gravitational theory.

The Maxwell field carries helicity, that is, spin projected along its
wave vector, but is doesn't carry spin proper as a gauge covariant
quantity. Therefore, there is no electromagnetic contribution to the
material spin on the right-hand-side of (\ref{seconda}) or
(\ref{secondb}). Light is insensitive to torsion; torsion cannot be
``seen''.\footnote{Only a nonminimal coupling of the electromagnetic
  field to torsion-square pieces is conceivable, see \cite{Itin:2003hr}.}

Torsion effects in EC-theory are minute. Besides the Einsteinian
gravitational field, we have additionally a very weak {\it spin-spin
  contact interaction\/} that is proportional to the gravitational
constant, which is measurable in principle. For a particle of mass
$m$ and reduced Compton wave length $\lambda_{\text{Co}}:=\hbar/mc$
(with $\hbar$ = reduced Planck constant, $c$ = speed of light), there
exists in EC a critical density and, equivalently, a critical radius
of ($\ell_{\text{P$\ell$}}$ = Planck length)
\begin{equation}\label{crit}
\rho_{\text{EC}}\sim
m/(\lambda_{\text{Co}}\ell_{\text{P$\ell$}}^2)\,\qquad\text{and}\qquad
r_{\text{EC}}\sim(\lambda_{\text{Co}}\ell_{\text{P$\ell$}}^2)^{1/3}\,,
\end{equation}
respectively, see \cite{RMP}. For a nucleon we have
$\rho_{\text{EC}}\approx 10^{54}\,$g/cm$^3$ and $r_{\text{EC}}\approx
10^{-26}\,$cm. Whereas those densities are extremely high {}from a
usual lab perspective or even {}from the point of view of a neutron
star ($\approx 10^{16}\,$g/cm$^3$), in cosmology they are standard. It
may be sufficient to recall that inflation is believed to set in
around the Planck density of $10^{93}\,$g/cm$^3$.

At densities higher than $\rho_{\text{EC}}$, EC-theory is expected to
overtake GR. There is no reason why GR should survive under those
conditions, since {\it for fermions} the gauge-theoretical framework
seems more trustworthy. Some cosmological models of EC can be found in
\cite{Reader}.

It is probably fair to say that EC has been established as a
consistent and viable theory of gravity and the Riemann--Cartan
geometry of spacetime has won solid support so that its study should
not be skipped in philosophical circles as undesirable complication of
the Riemann geometry of GR.

\section{Quadratic Poincar\'e gauge theory of gravity (qPG)}

Let me first express a word of caution: In a fairly recent paper,
{Mao, Tegmark, Guth, and Cabi} \cite{Mao} believe to have shown
``...that Gravity Probe B is an ideal experiment for further
constraining nonstandard torsion theories,...''  Nothing could be
further away {}from the truth. Following the guiding principle that
nothing is more practical than a good theory, Puetzfeld, Obukhov, et
al.\ \cite{Puetzfeld:2007ye,Hehl:2013} have shown that the measurement of
torsion requires elementary particle spins as test objects whereas in
Gravity Probe B the rotating quartz balls carry orbital angular
momentum only, but don't carry uncompensated {elementary particle
  spin}. Thus, the results in \cite{Mao} are simply incorrect in spite
of the wide publicity that this paper has won.

But back to Einstein--Cartan theory (EC). It is in many ways a very
degenerate theory. A contact interaction in physics cries for a
generalization to a propagating interaction, as has been the way
things developed in the Fermi theory of weak interaction---which was
a contact interaction par excellence---to the theory of the
propagating $W$ and $Z$. The recipe is very simple: The EC-Lagrangian
is linear in the Lorentz field strength, add terms that are quadratic
in the translational field strength (torsion) and the Lorentz field
strength (curvature).

Instead of boring you with all the details of this development to
quadratic Lagrangians in a RC-spacetime and who did what and when and
why, I shock you again with a messy formula. This is the most general
quadratic Lagrangian including parity violating pieces (see
\cite{Baekler:2011jt} and the explanations in the subsequent paragraphs):
\begin{eqnarray} \label{QMA} \hspace{-50pt} V &\!\!=\!\!&
  \,\frac{1}{2\kappa}[\,\left(
\,a_0R+{b_0}X
-2\Lambda)\,\eta\right.\\
\hspace{0pt} && \hspace{13pt}\left. +{\scriptstyle \frac{1}{3}} a_{2} {\cal
    V}\wedge {}^{\star\!} {\cal V} -{\scriptstyle \frac{1}{3}}{a_3}{{\cal A}
    \wedge{} ^{\star\!\!\!}{\cal A}} -{\scriptstyle \frac{2}{3}}\sigma_{2} {\cal
    V}\wedge{} ^{\star\!\!\!} {\cal A}+ a_{1}{}^{(1)}T^\alpha
  \wedge {}^{\star(1)}T_\a \right]\nonumber \\
\hspace{0pt} & &\hspace{-11pt} -\frac{1}{2\varrho} \left[
  ({\scriptstyle\frac{1}{12}}w_6 R^2- {\scriptstyle\frac{1}{12}}
w_3 {X^2} +{\scriptstyle    \frac{1}{12}}
  \mu_3 R
  X)\,\eta + w_{4}{}^{(4)}\!R^{\a\beta}\wedge
  {}^{\star(4)}\!R_{\a\beta}\right.\nonumber\cr \hspace{-50pt} && \\
\hspace{0pt} & &\left. +{}^{(2)}\!R^{\a\beta}\wedge (
  w_2{}^{\star(2)}\!R_{\a\beta}
  +\mu_2{}^{(4)}\!R_{\a\beta})+{}^{(5)}\!R^{\a\beta}
  \wedge(w_5{}^{\star(5)}\!R_{\a\beta}
  +\mu_4{}^{(5)}\!R_{\a\beta})\right].\nonumber
\end{eqnarray}
The first two lines represent {\it weak} gravity, with the
conventional gravitational constant $\kappa$, the last two lines
speculative {\it strong} gravity with the dimensionless strong gravity
constant $\varrho$. The unknown constants $(a_0;a_1,a_2,a_3;b_0,\sigma_2)$,
weight the different terms of weak gravity, the unknown constants
$(w_2,w_3,w_4,w_5,w_6;\mu_2,\mu_3,\mu_4)$ those of strong
gravity. What a mess!

But let us discuss the formula line by line: In the {\it first line}
$R$ is the EC-term, $X:=\frac
14\eta_{\alpha\beta\gamma\delta}R^{[\alpha\beta\gamma\delta]}$ is the
(parity violating) curvature pseudoscalar, which vanishes in
Riemannian space, but is nonvanishing in RC-space. This term is
presently very popular in the quantum gravity scene, $\Lambda$ is the
cosmological constant, and $\eta$ the `volume element'.

The {\it second line} houses all torsion-square pieces. We have a
tensor torsion $^{(1)}T^\alpha$, a vector torsion $\cal V$ and an
axial vector torsion $\cal A$. They can enter in the combinations
shown. The remarkable fact is that for dimensional reasons the first
line and the second line give rise to similar effects. Instead of the
EC-theory with $R$, you can select a suitable linear combination of
torsion-square pieces acting in a RC-space with vanishing curvature
(Weitzenb\"ock space), see, for example, Itin \cite{YakovTele} or
\cite{Reader} and the historical article of Sauer \cite{Sauer}. On the
first two lines there are literally hundreds of published papers
studying different properties. Numerous printed pages could be saved,
if our colleagues would start with the first two lines right away and
just motivate their choice of the unknown constants.

Now we turn to the remaining more speculative pieces, which are,
however, fairly plausible due to their Yang--Mills type
structure. After all, C.~N.~Yang himself proposed such a theory
\cite{Yang:1974}. We are not in bad company! In the {\it third line}
we turn our attention immediately to the first three pieces: They are
just squares built {}from the curvature scalar and/or the curvature
pseudoscalar. The curvature in a RC-space $R^{\a\b}$ decomposes into 6
irreducible pieces $^{(I)}\!R^{\a\b}$: they are numbered by $I$, running
{}from 1 to 6. The pseudoscalar $X$ is number 3, the scalar $R$ number
6. The last term in the third line is then a square piece of number
4. In the {\it fourth line} we have the remaining curvature square
pieces. The term with number 1 drops out due to certain identities.

This is only algebra. Where is the physics? you may ask. Well, we have
to find out. It will be a task of the future to single out of this set
of quadratic Lagrangians (\ref{QMA}) the physically acceptable one.
How such possible developments may look like, I will illustrate with
one example. Shie--Nester--Yo \cite{Shie:2008ms} developed a fairly
realistic cosmological model of Friedmann type with {\it propagating
  connection} by picking the Lagrangian
\begin{equation}\label{SNY}
  V_{\rm SNY}=\underbrace{\frac{1}{2\kappa}\left(a_0R\,\eta+ a_{1}{}^{(1)}T^\alpha
      \wedge {}^{\star(1)}T_\a \right)}_{\text{weak Newton-Einstein
      gravity}} \hspace{15pt} -\underbrace{\frac{w_6}{24\varrho} 
    R^2\,\eta}_{\text{strong YM-type gravity}}\,.
\end{equation}
They found two conventional graviton helicities, as in GR, and this,
for $a_0\ne a_1$, combined with a torsion mode of mass of
$\mu:=a_1-a_0$ and spin $0^+$ (spin zero with positive
parity, that is, an ordinary scalar), which has many attractive features. Of
course, equation (\ref{SNY}) is a subcase of equation (\ref{QMA}). In
the meantime this paper has been generalized by including parity
violating pieces, inter alia, and it has been numerically
evaluated. This paper has about 45 follow-up papers. In this way one
collects more and more insight into the possible physics behind the
most general quadratic PG-Lagrangian.

\section{Outlook}

What is the benefit of all of that for the theory of spacetime? Well,
it is a small but decisive step beyond the established Riemannian
spacetime structure of GR. Cartan's torsion has been incorporated into
the body of knowledge of classical spacetime geometry. At the same
time it has been demonstrated that the Poincar\'e group
$P(1,3)=T(4)\!\rtimes\! SO(1,3)$, acting in the Minkowski space, and
the behavior of the Dirac field in non-inertial frames leads, via the
gauge principle, to the Riemann--Cartan geometry of spacetime. That
is, the $P(1,3)$ symmetry induced the Riemann--Cartan geometry.

The generalization of this procedure seems to be straightforward. If
we add the group of dilations to $P(1,3)$, assuming scale covariance
in addition to the $P(1,3)$ covariance, we arrive at the 11 parametric
Weyl group. Gauging it, requires one more potential, namely the Weyl
covector $Q$, defined in terms of the nonmetricity according to
$Q:=\frac 14 Q_\g{}^\g=\frac 14 g^{\a\b}Q_{\a\b}$, see equation
(\ref{nonm}). Associated with it comes a conservation law and the
Noether current $\Delta=\d L/\d Q$, the dilation or scale current,
which Weyl had mistaken for the electric current. If we turn the
crank, a Weyl--Cartan spacetime emerges together with a gauge field
equation that has the dilation current as source. This is standard Weyl
lore from a contemporary point of view, see \cite{Reader}, Chapter 8.

I hope it doesn't take you by surprise that I cannot see much common
ground with the theory of E.~Scholz presented during this workshop
\cite{Scholz}. In his approach, spacetime is governed by a Weyl
geometry with vanishing torsion, but the dilation current is not an
inhabitant of the Weyl space of Scholz---or, at least, this current
has not been identified as such and lives anonymously and drifts
around uncontrolled by any field equation.

Instead, one can add to the $P(1,3)$ simple supersymmetry (symmetry
between fermions and bosons) by extending the Poincar\'e algebra with
anticommuting fermionic generators thus being led to a Poincar\'e
superalgebra. The corresponding gauge procedure creates a so-called
superspace(time) geometry. The field equations of simple supergravity
can be immediately written down by using the EC-field equations
(\ref{firstb}) and (\ref{secondb}); as sources one takes the
energy-momentum and the spin currents of the massless
Rarita--Schwinger field, which carries spin $\frac 32$. The
Rarita--Schwinger field conspires with the effective spin 2 of the
EC-field to build up a super multiplet $(2,\frac 32)$, compare
\cite{Reader}, Chapter 12.

In this way we see that also in supersymmetry the gauge concept of
Weyl and Yang--Mills--Utiyama is successful. And the geometry of
spacetime turned out to have a potential ``super'' structure beyond
Riemann--Cartan geometry.

Mielke \cite{Mielke:2011zza} generalized the Poincar\'e group
$T(4)\!\rtimes\! SO(1,3)$ to the $SL(5,R)$ and recovered by symmetry
breaking reasonable 4-dimensional gravitational gauge structures. This
could be a future-pointing approach.

\begin{footnotesize}

\subsection*{Acknowledgments} 

I would like to thank Dennis Lehmkuhl, Erhard Scholz, and Gregor
Schiemann most sincerely for the invitation to this workshop, for the
hospitality, and for the lively discussions. I am also grateful to
Peter Baekler (D\"usseldorf), Milutin Blagojevi\'c (Belgrade), Yakov
Itin (Jerusalem), Claus Kiefer (Cologne), Bahram Mashhoon (Columbia,
Missouri), Eckehard Mielke (Mexico City), James Nester (Chung-li),
Yuri Obukhov (Co\-logne/Moscow), and Dirk Puetzfeld (Bremen) for many
oral or email discussion on the subject of gravity and gauging. The
criticism of two referees also helped to improve this article. I am
particularly grateful to J.~Brian Pitts (Cambridge, UK) for detailed
remarks and suggestions. This work was supported by the German-Israeli
Foundation for Scientific Research and Development (GIF), Research
Grant No.\ 1078-107.14/2009.

\end{footnotesize}

\centerline{=======}
\end{document}